# Dispersion forces in the Lifshitz problem

M. V. Davidovich[1]

[1]Saratov National Research State University named after N. G. Chernyshevsky,

410012 Saratov, Russia

davidovichmv@info.sgu.ru

On the basis of the mode matching technique, the surface density of the dispersion force between the half-spaces in the Lifshitz problem is sequentially derived. An expression of the force density is obtained using the fluctuation-dissipation theorem and the energy-momentum tensor in vacuum. A new approach to calculating the Casimir-Lifshitz force per unit area based on the Green's function method of classical electrodynamics and the Lorentz force is proposed. The result is presented as a double integral over the real frequency axis and the real axis in the pulse space. It is shown that the origin of the force is associated with the existence of fluctuating slow coupled long-wave surface polaritons.

Fluctuating electromagnetic fields are responsible for such important physical phenomena as thermal radiation, radiative heat transfer, Van der Waals interactions at short distances between molecules and nanoclusters, including Van der Waals friction, the Casimir effect, and the Casimir-Lifshitz forces between bodies [1–15]. The interaction between micro-and nano-objects began to be widely studied with the work of Casimir [6]. It can have the character of attraction or repulsion. The Lifshitz theory [8] for an arbitrary temperature was constructed for objects extended in depth, separated by a flat vacuum gap. Extension means the attenuation of radiation deep into the structure. It determines the force based on the classical energy-momentum tensor (EMT) in a vacuum using the fluctuation-dissipation theorem. In a number of papers (see [12–14]), the theory of dispersion forces is developed on the basis of the quantum-field approach in statistical physics with the use of Green's temperature functions (FG). However, the calculation of forces for complex configurations of structures based on this approach is very complex.

In this paper, we consider in detail the derivation of the Casimir-Lifshitz dispersion forces of the Lifshitz structure in the form of a vacuum gap between two half-spaces, which for simplicity are considered the same and dielectric. The Lifshitz method is used in the form of the decomposition of fields in non-homogeneous plane waves (TM and TE). The Lorentz model and the Drude-Smith model, which describe metals well, are used for the permittivity. To determine the spectral correlation relations, two dielectric layers of finite thickness $t$ with the transition of this thickness to the infinite limit are also considered. An approach to the solution of the same problem is also outlined, it is based not on the decomposition of fields, but on their representation through the classical tensor GF of electrodynamics. The determination of the force density is performed in terms of the Maxwell tension tensor (MTT), the form of which is known for vacuum. In addition, a microscopic approach to the determination of the force based on the Maxwell-Lorentz equations and the formula for the Lorentz force is outlined, which can also be applied to interacting bodies in dispersing media.

Let us consider two parallel rectangular dielectric layers with thicknesses and infinite along the $x$ and $y$ and located at a distance $2a$ along $z$ axes (Fig. 1). The layers are located in the $-b \leq z \leq -a$ and $a \leq z \leq b$ areas. The vacuum gap between them is $d = 2a$. There are polarization currents $\mathbf{J}^P = \partial_t \mathbf{P}$ in the layers $\mathbf{J}^P(\omega, \mathbf{r}) = i\omega\varepsilon_0(\varepsilon(\omega, \mathbf{r}) - 1)\mathbf{E}(\omega, \mathbf{r})$. They excite a secondary electromagnetic field, and a self-consistent system of equations is generally a system of volume integral equations (for conducting surfaces, these are surface integral equations). Next, we denote $\mathbf{J}^P = \mathbf{J}$. In the considered structure, free waves (without excitation by external sources) – plasmon-polaritons (PP) – are possible. We consider the forced waves caused by fluctuation fluctuations. The current and field densities in each of the layers (bodies) and fields will be denoted by $\mathbf{J}_n(\omega, \mathbf{r})$, $\mathbf{E}_n(\omega, \mathbf{r})$, $\mathbf{H}_n(\omega, \mathbf{r})$ ($n=1,2$), and the index "0" – the corresponding fluctuation values, for example, $\mathbf{J}_n^0$. The quantities $\sigma_n = i\omega\varepsilon_0(\varepsilon_n(\omega) - 1) = Z_0^{-1}k_0\xi_n$ can be considered as conductivities, since $\mathbf{J}_n(\omega, \mathbf{r}) = \sigma_n(\omega)\mathbf{E}_n(\omega, \mathbf{r})$. Here $Z_0 = \sqrt{\mu_0/\varepsilon_0}$ and the dimensionless quantities are introduced $\xi_n = i(\varepsilon_n(\omega) - 1)$. We will describe the dispersion of the permittivity (DP) by the Lorentz formula. It can have several resonant frequencies corresponding to the transition frequencies in the excitation spectra of atoms. We will consider several transition frequencies $\omega_{0k}$, considering the rest to be substantially higher, and their contribution at lower frequencies to be constant and described by a constant term $\varepsilon_L > 1$. When calculating the force, we will limit the range to the specified frequencies. Then we have

$$\varepsilon(\omega) = \varepsilon'(\omega) - i\varepsilon''(\omega) = \varepsilon_L + \sum_{k=1}^{K} \frac{\omega_{pk}^2}{\omega_{0k}^2 - \omega(\omega - i\omega_{ck})}. \qquad (1)$$

The square of the "plasma frequency" $\omega_{pk}^2$ is related to the strength of the oscillator of the corresponding transition and the concentration of atoms. The concentration of atoms is considered constant, so the DP does not depend on the coordinates. This is the simplest case for analyzing the dispersion forces. For real substances, this formula can be used as an approximation, taking into account the significant dependence of the DP on the internal field, temperature, and other factors. Then the real dielectric can be approximated by contributions from several resonant transition frequencies. Next, we will also describe the metal with a similar formula, supplementing it with the Drude-Smith term. What matters to us is that, as well as the frequency behavior :

В общем случае считаем ДП различными, т.е. вводим зависящие от *n* величины. Далее металл также будем описывать подобной формулой, дополняя ее членом Друде-Смита. Нам важно, что $\varepsilon''(0) = 0$, а также частотное поведение $\varepsilon''(\omega) = -\text{Im}(\varepsilon''(\omega))$:

$$\varepsilon''(\omega) = \sum_n \frac{\omega_{np}^2 \omega \omega_{nc}}{\left(\omega_{n0}^2 - \omega^2\right)^2 + \left(\omega \omega_{nc}\right)^2}.$$

Next, we consider the case $t \to \infty$, Fig. 1. We write the Maxwell spectral equations (we omit the frequency $\omega$) in the form,

$$\nabla \times \mathbf{H}(\mathbf{r}) = ik_0 Z_0^{-1} \varepsilon(\mathbf{r}) \mathbf{E} + \mathbf{J}^0(\mathbf{r}), \quad \nabla \times \mathbf{E}(\mathbf{r}) = -ik_0 Z_0 \mathbf{H}(\mathbf{r}). \qquad (2)$$

We have introduced the fluctuation current density into these equations $\mathbf{J}^0$, so they are inhomogeneous. The dependence on the coordinates of the DP is understood as follows: in the region of half-spaces it is constant, and in the region of the gap $\varepsilon = 1$. In such a dielectric, a polarization current $\partial_t \mathbf{P}(\mathbf{r}) = i\omega \mathbf{P}(\mathbf{r})$ with a density occurs $\mathbf{J}_p(\mathbf{r}) = i\omega \varepsilon_0 (\varepsilon(\mathbf{r}) - 1) \mathbf{E}(\mathbf{r})$. It is missing in the gap. Equations (2) are rewritten in the form

$$\nabla \times \mathbf{H}(\mathbf{r}) = ik_0 Z_0^{-1} \mathbf{E} + \mathbf{J}^0(\mathbf{r}) + \mathbf{J}_p(\mathbf{r}), \quad \nabla \times \mathbf{E}(\mathbf{r}) = -ik_0 Z_0 \mathbf{H}(\mathbf{r}). \qquad (3)$$

He shows that the equations can be considered in a vacuum by introducing an additional polarization current density. The word "density" will often be omitted further, meaning by the word current its corresponding density. The fluctuation current is not present in the gap, so it can also be represented in the form $\mathbf{J}^0(\mathbf{r}) = i\omega \varepsilon_0 (\varepsilon(\mathbf{r}) - 1) \mathbf{E}^0(\mathbf{r})$ where a certain associated fluctuation field is introduced. It is defined only in the dielectric. It is possible to consider a fluctuating current as the sum of the currents in the two half-spaces $\mathbf{J}^0 = \mathbf{J}_1^0 + \mathbf{J}_2^0$. The dispersion in the media is related to the movement of charges, i.e., to the current. The power of the fluctuations is determined for $\mathbf{J}^0(\mathbf{r})$ and $\mathbf{J}^0(\mathbf{r})$ by $\varepsilon''(\mathbf{r})$. In this case, the medium can be conductive, or not. In the theory of dispersion, there is no difference between the conduction current and the polarization current associated with charge fluctuations near the atoms: both are described by the polarization and the polarization current, which includes the conduction current. In a conductive medium is free charges, and it is often described by the Drude dispersion of Lorenza. Such a variance is obtained if one of the frequencies is set to zero in (1). Then in (1) there is a term (we took $\omega_{00}^2 = 0$):

$$\frac{-\omega_{p0}^2}{\omega(\omega - i\omega_{ck})}.$$

This term has a pole at zero, which leads to a number of problems in the theory of dispersion forces, in particular, to a violation of the third principle of thermodynamics – the Nernst thermal theorem (see [15]). Equality $\omega_{00}^2 = 0$ means the separation of an electron from an atom (an infinitely weak bond).

However, in finite structures and clusters, the electron cannot go to infinity, and the influence of the boundaries leads to the fact that it is somehow weakly bound, which can be described by a finite value $\omega_{00}^2 = (c\pi/L)^2$. Here $L$ is the size of the structure. This corresponds to the Drude-Smith model [16, 17], which allows us to avoid a number of difficulties (divergence of integrals, violation of the Nernst thermal theorem, etc.).

In [18] (formulas (4.8), (4.9)), on the basis of the fluctuation-dissipation theorem, correlation relations for fluctuation currents in a conducting medium in the Gauss system are obtained for an equilibrium medium (see also [19]). The dissipation in the conducting medium is taken into account as the conductivity, which leads to the appearance of the imaginary part of the DP. The result is also valid for a nonconducting dielectric, in which the dissipation is described by the imaginary part of the DP. With the time dependence used, this imaginary part is negative. To switch to SI, we multiply the result [18] by $4\pi\varepsilon_0$:

$$\left\langle J_\nu^0(\mathbf{r}), J_\mu^0(\mathbf{r}') \right\rangle = \omega^2 \varepsilon_0^2 \left\langle E_\nu^0(\mathbf{r}), E_\nu^0(\mathbf{r}') \right\rangle = \delta_{\nu\mu} \delta(\mathbf{r}-\mathbf{r}') F(\omega,T), \tag{4}$$

$$F(\omega,T) = \omega \varepsilon_0 \varepsilon'' \Theta(\omega,T)/\pi, \tag{5}$$

$$\Theta(\omega,T) = \frac{\hbar\omega}{2} + \frac{\hbar\omega}{\exp(\hbar\omega/k_B T)-1} = \frac{\hbar\omega}{2} \coth\left(\frac{\hbar\omega}{2k_B T}\right). \tag{6}$$

Here the values $\nu$, $\mu$ run through of $x$, $y$, and $z$, and the function (6) gives the average energy of the quantum oscillator at temperature $T$ [18]. To perform (4), it is necessary and sufficient for the spectral correlations to be performed $\left\langle J_{n\nu}^0(\mathbf{k}), J_{m\mu}^0(\mathbf{k}') \right\rangle = (2\pi)^3 \delta_{nm} \delta_{\nu\mu} \delta(\mathbf{k}-\mathbf{k}') F(\omega,T)$. Here we denote the spectral densities in different half-spaces ($n=1,2$). In (5), this independence of correlations is provided by the delta function, since the points in different half-spaces are always different. However, the condition of the delta-correlation of the spectral components is true only in the case of an infinite medium. For the half-spaces in question, this is not the case. It is approximately made for a narrow gap. We have

$$J_{n\nu}^0(\mathbf{r}) = \frac{1}{(2\pi)^3} \int_{-\infty}^{\infty} \exp(i\mathbf{kr}) J_{n\nu}^0(\mathbf{k}) d^3k,$$

$$J_{n\nu}^0(\mathbf{k}) = J_{n\nu}^0(\mathbf{q},\gamma) = \int_{-\infty}^{\infty} \exp(i\mathbf{qr}_\tau) \left[ \int_{-\infty}^{\infty} \exp(i\gamma z) J_{n\nu}^0(\mathbf{r}_\tau,z) dz - \int_{-a}^{a} \exp(i\gamma z) J_{n\nu}^0(\mathbf{r}_\tau,z) dz \right] d^2r.$$

Here $\mathbf{r}_\tau = (x,y)$, $\mathbf{q} = (k_x,k_y)$, $\mathbf{k} = (k_x,k_y,\gamma)$. Calculating the correlations, we find

$$\left\langle J_{n\nu}^0(\mathbf{q},\gamma), J_{m\mu}^0(\mathbf{q}',\gamma') \right\rangle = (2\pi)^3 \delta_{nm} \delta_{\nu\mu} \delta(\mathbf{q}-\mathbf{q}') F(\omega,T) \tilde{F}(\gamma,\gamma'), \tag{7}$$

$$\left\langle J_{n\nu}^0(\mathbf{q},\gamma), J_{m\mu}^0(\mathbf{q},\gamma') \right\rangle = (2\pi)^3 \delta_{nm} \delta_{\nu\mu} F(\omega,T) \tilde{F}(\gamma,\gamma'),$$

$$\tilde{F}(\gamma,\gamma')=\delta(\gamma-\gamma')-(a/\pi)\mathrm{sinc}((\gamma-\gamma')a). \tag{8}$$

The function $\mathrm{sinc}(x)=\sin(x)/x$ is indicated here. The relation (7) implies a delta-correlation in the absence of a gap $a=0$. In the case of an infinitely wide gap $a\to\infty$, we have $2a\,\mathrm{sinc}((\gamma-\gamma')a)\to 2\pi\delta(\gamma-\gamma')$, and (8) gives zero. This is a condition for the absence of a structure. We have $\langle E_\nu^0(\mathbf{r}),E_\nu^0(\mathbf{r})\rangle = \langle j_\nu^0(\mathbf{r}), j_\nu^0(\mathbf{r})\rangle/(\omega\varepsilon_0)^2 = \Theta(\omega,T)\varepsilon''/(\pi\omega\varepsilon_0)$. Writing the inverse Fourier transform

$$\mathbf{J}^0(\mathbf{r})=\frac{1}{(2\pi)^3}\int_{-\infty}^{\infty}\exp(-i\mathbf{kr})\mathbf{J}^0(\mathbf{k})d^3k,$$

$$\mathbf{E}^0(\mathbf{r})=\frac{1}{(2\pi)^3}\int_{-\infty}^{\infty}\exp(-i\mathbf{kr})\mathbf{E}^0(\mathbf{k})d^3k,$$

and using (2) we have

$$\mathbf{H}^0(\mathbf{r})=\frac{1}{(2\pi)^3 k_0^2}\int_{-\infty}^{\infty}\exp(-i\mathbf{kr})\hat{R}(\mathbf{k})\mathbf{E}^0(\mathbf{k})d^3k,$$

The matrix operator is denoted here

$$\hat{R}(\mathbf{k})=i\begin{bmatrix} 0 & \gamma & -k_y \\ -\gamma & 0 & k_x \\ k_y & -k_x & 0 \end{bmatrix},$$

performing the operation of taking the rotor. We have specifically separately designated the $z$-component of the wave vector as $\gamma$, since we introduce the notation $k_z=\sqrt{k_0^2\varepsilon-k_x^2-k_y^2}$, $k_z^0=\sqrt{k_0^2-k_x^2-k_y^2}$. These are the constant propagation of free inhomogeneous plane waves along the $z$-axis in the dielectric and in the vacuum. Such waves are classified as E-waves and H-waves with normalized (dimensionless) wave resistances $\rho_e=k_z/(\varepsilon k_0)=\sqrt{k_0^2\varepsilon-k_x^2-k_y^2}/(\varepsilon k_0)$, $\rho_h=k_0/k_z=k_0/\sqrt{k_0^2\varepsilon-k_x^2-k_y^2}$ in a dielectric, and $\rho_e^0=k_z/k_0=\sqrt{k_0^2-k_x^2-k_y^2}/k_0$, $\rho_h^0=k_0/k_z=k_0/\sqrt{k_0^2-k_x^2-k_y^2}$ in a vacuum. For brevity, we denote $\mathbf{r}_\tau=(x,y)$, $e_{\mathbf{q}}(\mathbf{r}_\tau)=\exp(-i\mathbf{qr}_\tau)$. The vector $\mathbf{q}$ defines the propagation angles of inhomogeneous plane waves. By values, its component should be integrated to find the complete fields. Accordingly, the fields in the coordinate representation have the form

$$\mathbf{E}(\mathbf{r}_\tau,z)=\frac{1}{(2\pi)^2}\int_{-\infty}^{\infty}e_{\mathbf{q}}(\mathbf{r}_\tau)\mathbf{E}(\mathbf{r}_\tau,z)d^2q,$$

and similarly for the magnetic field.

The transverse fields are expressed by the formulas (10.53) and (10.54) of [20] (see also [21, 22]) in terms of the longitudinal $z$-components in the vacuum gap $E_{0z}(\mathbf{q},\mathbf{r}_\tau,z)=e_\mathbf{q}(\mathbf{r}_\tau)E_{0z}(\mathbf{q},z)$, $H_{0z}(\mathbf{q},\mathbf{r}_\tau,z)=e_\mathbf{q}(\mathbf{r}_\tau)H_{0z}(\mathbf{q},z)$, where

$$E_{0z}(\mathbf{q},z)=E_{0z}(\mathbf{q})\exp(-ik_z^0(z+a))-E_{0z}(\mathbf{q})\exp(ik_z^0(z-a)), \tag{9}$$

$$H_{0z}(\mathbf{q},z)=H_{0z}(\mathbf{q})\exp(-ik_z^0(z+a))+H_{0z}(\mathbf{q})\exp(ik_z^0(z-a)), \tag{10}$$

and in the dielectric $E_z(\mathbf{q},\mathbf{r}_\tau,z)=e_\mathbf{q}(\mathbf{r}_\tau)E_z(\mathbf{q},z)$, $H_z(\mathbf{q},\mathbf{r}_\tau,z)=e_\mathbf{q}(\mathbf{r}_\tau)H_z(\mathbf{q},z)$, in which

$$E_z(\mathbf{q},z)=E_z(\mathbf{q})\exp(-ik_z|z+a|)+E_z(\mathbf{q})\exp(-ik_z|z-a|), \tag{11}$$

$$H_z(\mathbf{q},z)=H_z(\mathbf{q})\exp(-ik_z|z+a|)+H_z(\mathbf{q})\exp(-ik_z|z-a|). \tag{12}$$

In the gap, we write the fields as waves moving in two directions of the axis $\pm z$. It will be shown later that only evanescent waves should be taken into account, for which the constant $k_z^0$ is imaginary, i.e., take $\kappa^0=ik_z^0=\sqrt{q^2-k_0^2}$, $q^2=q_x^2+q_y^2=k_x^2+k_y^2$. In the dielectric, the fields are written as waves diverging from the boundaries of the gap. The transverse components of the waves will be expressed in terms of the transverse amplitudes of the electric field. They are expressed in terms of the longitudinal as $E_{0x}^e(\mathbf{q})=-k_z^0 k_x q^{-2}E_{0z}(\mathbf{q})$, $E_{0y}^e(\mathbf{q})=-k_z^0 k_y q^{-2}E_{0z}(\mathbf{q})$, $E_x^e(\mathbf{q})=-k_z k_x q^{-2}E_z(\mathbf{q})$, $E_y^e(\mathbf{q})=-k_z k_y q^{-2}E_z(\mathbf{q})$. Also, the longitudinal components of the magnetic field, expressed in terms of the transverse components of the H-waves of the electric field, will be required further. They are given by the connections $E_{0x}^h(\mathbf{q})=-k_z^0 k_y Z_0 \rho_0^h H_{0z}(\mathbf{q})/q^2$, $E_{0y}^h(\mathbf{q})=k_z^0 k_x Z_0 \rho_0^h H_{0z}(\mathbf{q})/q^2$, $E_x^h(\mathbf{q})=-k_z k_y Z_0 \rho^h H_z(\mathbf{q})/q^2$, $E_y^h(\mathbf{q})=k_z k_x Z_0 \rho^h H_z(\mathbf{q})/q^2$. In addition, you should write the expressions of the transverse components of the magnetic field through its longitudinal components: $H_{0x}^h(\mathbf{q})=-k_z^0 k_x H_{0z}(\mathbf{q})/q^2$, $H_{0y}^h(\mathbf{q})=-k_z^0 k_y H_{0z}(\mathbf{q})/q^2$.

The transverse components of the E-waves in the gap have the form

$$E_{0x}^e(\mathbf{q},z)=E_{0x}^e(\mathbf{q})\exp(-ik_z^0(z+a))+E_{0x}^e(\mathbf{q})\exp(ik_z^0(z-a)),$$

$$H_{0y}^e(\mathbf{q},z)=\frac{E_{0x}^e(\mathbf{q})}{Z_0\rho_0^e}\exp(-ik_z^0(z+a))-\frac{E_{0x}^e(\mathbf{q})}{Z_0\rho_0^e}\exp(ik_z^0(z-a)),$$

$$E_{0y}^e(\mathbf{q},z)=E_{0y}^e(\mathbf{q})\exp(-ik_z^0(z+a))+E_{0y}^e(\mathbf{q})\exp(ik_z^0(z-a)),$$

$$H_{0x}^e(\mathbf{q},z)=-\frac{E_{0y}^e(\mathbf{q})}{Z_0\rho_0^e}\exp(-ik_z^0(z+a))+\frac{E_{0y}^e(\mathbf{q})}{Z_0\rho_0^e}\exp(ik_z^0(z-a)).$$

The transverse components of the H-waves in the gap have the form

$$E_{0x}^h(\mathbf{q},z)=E_{0x}^h(\mathbf{q})\exp(-ik_z^0(z+a))+E_{0x}^h(\mathbf{q})\exp(ik_z^0(z-a)),$$

$$H_{0y}^{h}(\mathbf{q},z) = \frac{E_{0x}^{h}(\mathbf{q})}{Z_0 \rho_0^{h}} \exp(-ik_z^0(z+a)) - \frac{E_{0x}^{h}(\mathbf{q})}{Z_0 \rho_0^{h}} \exp(ik_z^0(z-a)),$$

$$E_{0y}^{h}(\mathbf{q},z) = E_{0y}^{h}(\mathbf{q})\exp(-ik_z^0(z+a)) + E_{0y}^{h}(\mathbf{q})\exp(ik_z^0(z-a)),$$

$$H_{0x}^{h}(\mathbf{q},z) = -\frac{E_{0y}^{h}(\mathbf{q})}{Z_0 \rho_0^{h}} \exp(-ik_z^0(z+a)) + \frac{E_{0y}^{h}(\mathbf{q})}{Z_0 \rho_0^{h}} \exp(ik_z^0(z-a)).$$

The transverse components of free E-waves in a dielectric have the form

$$E_x^e(\mathbf{q},z) = E_x^e(\mathbf{q})\exp(-ik_z|z+a|) + E_x^e(\mathbf{q})\exp(-ik_z|z-a|),$$

$$H_y^e(\mathbf{q},z) = \frac{E_x^e(\mathbf{q})}{Z_0 \rho^e} \exp(-ik_z|z+a|) - \frac{E_x^e(\mathbf{q})}{Z_0 \rho^e} \exp(-ik_z|z-a|),$$

$$E_y^e(\mathbf{q},z) = E_y^e(\mathbf{q})\exp(-ik_z|z+a|) + E_y^e(\mathbf{q})\exp(-ik_z|z-a|),$$

$$H_x^e(\mathbf{q},z) = -\frac{E_y^e(\mathbf{q})}{Z_0 \rho^e} \exp(-ik_z|z+a|) + \frac{E_y^e(\mathbf{q})}{Z_0 \rho^e} \exp(-ik_z|z-a|).$$

The transverse components of free H-waves in a dielectric are written as

$$E_x^h(\mathbf{q},z) = E_x^h(\mathbf{q})\exp(-ik_z|z+a|) + E_x^h(\mathbf{q})\exp(-ik_z|z-a|),$$

$$H_y^h(\mathbf{q},z) = \frac{E_x^h(\mathbf{q})}{Z_0 \rho^h} \exp(-ik_z|z+a|) - \frac{E_x^h(\mathbf{q})}{Z_0 \rho^h} \exp(-ik_z|z-a|),$$

$$E_y^h(\mathbf{q},z) = E_y^h(\mathbf{q})\exp(-ik_z|z+a|) + E_y^h(\mathbf{q})\exp(-ik_z|z-a|),$$

$$H_x^h(\mathbf{q},z) = -\frac{E_y^h(\mathbf{q})}{Z_0 \rho^h} \exp(-ik_z|z+a|) + \frac{E_y^h(\mathbf{q})}{Z_0 \rho^h} \exp(-ik_z|z-a|).$$

In addition to these fields, the components of the fluctuating electric field $\mathbf{E}^0$ and the magnetic field $\mathbf{H}^0(\mathbf{r}) = i(k_0 Z_0)^{-1} \nabla \times \mathbf{E}^0(\mathbf{r})$ should be recorded. After converting to **q**-space

$$E_\nu^0(\mathbf{q},z) = \frac{1}{2\pi} \int_{-\infty}^{\infty} \exp(-i\gamma z) E_\nu^0(\mathbf{q},\gamma) d\gamma,$$

$$H_x^0(\mathbf{q},z) = \frac{1}{2\pi k_0 Z_0} \int_{-\infty}^{\infty} \exp(-i\gamma z)[k_y E_z^0(\mathbf{q},\gamma) - \gamma E_y^0(\mathbf{q},\gamma)] d\gamma,$$

$$H_y^0(\mathbf{q},z) = \frac{1}{2\pi Z_0} \int_{-\infty}^{\infty} \exp(-i\gamma z)[\gamma E_x^0(\mathbf{q},\gamma) - k_x E_z^0(\mathbf{q},\gamma)] d\gamma.$$

These fields have the form of plane waves moving in different directions. Assume that the thickness of the half-spaces is finite and equal to *t*. Then we go to the limit $t \to \infty$. Assuming that $E_\nu^0(\mathbf{q},z) = E_\nu^0(\mathbf{q})$, i.e., the spectral densities are independent of *z*, we obtain

$$E_\nu^0(\mathbf{q},\gamma) = \int_{a<|z|<a+t} \exp(-i\gamma z) E_\nu^0(\mathbf{q}) dz = E_\nu^0(\mathbf{q}) \int_{a<|z|<a+t|} \cos(\gamma z) dz = 2a E_\nu^0(\mathbf{q})\mathrm{sinc}(\gamma a).$$

Here, when moving to the limit, we took into account that the medium is dissipative, and at its large thickness, the waves are attenuated at $z = \pm t$. We will stitch the fields at $z = a$. The fields will be stitched together at $z = -a$ automatically due to the symmetry. Then

$$E_\nu^0(\mathbf{q},a) = (1/\pi)E_\nu^0(\mathbf{q})\int_0^\infty \frac{\sin(2a\gamma)}{\gamma}d\gamma = E_\nu^0(\mathbf{q})/2.$$

Similarly

$$H_x^0(\mathbf{q},a) = \frac{E_z^0(\mathbf{q})k_y}{2k_0 Z_0}$$

$$H_y^0(\mathbf{q},a) = \frac{-k_x E_z^0(\mathbf{q})}{2k_0 Z_0}.$$

The two denominators here appear due to the fact that only the intensities of the right half-plane contribute, and the spectral amplitudes do not depend on $a$. Since, then for the correlations $\langle E_\nu^0(\mathbf{q},z), E_\nu^0(\mathbf{q}',z')\rangle = (2\pi)^2 \delta(\mathbf{q}-\mathbf{q}'^2)\delta(z-z')F(\omega,T)/(\omega\varepsilon_0)^2$ at the crosslinking point, we have

$$\langle E_\nu^0(\mathbf{q},a), E_\nu^0(\mathbf{q}',a)\rangle = \langle E_\nu^0(\mathbf{q},-a), E_\nu^0(\mathbf{q}',-a)\rangle = (2\pi)^2 \delta(\mathbf{q}-\mathbf{q}')F(\omega,T)/(2\omega\varepsilon_0)^2. \quad (13)$$

We find correlations of spectral random fields. It should be noted that although they are not fully delta-correlated, i.e. the function (8) takes place, the result takes the form:

$$\langle E_\nu^0(\mathbf{q},\gamma), E_\mu^0(\mathbf{q}',\gamma')\rangle = (2\pi)^3 \delta_{\nu\mu}\delta(\mathbf{k}-\mathbf{k}')\frac{Z_0^2 F(\omega,T)}{k_0^2}, \quad (14)$$

$$\langle H_x^0(\mathbf{q},\gamma), H_x^0(\mathbf{q}',\gamma')\rangle = (2\pi)^3 \delta(\mathbf{k}-\mathbf{k}')\frac{\gamma\gamma' + k_y k_y'}{k_0^4}F(\omega,T). \quad (15)$$

It is the same as when $\tilde{F}(\gamma,\gamma') = \delta(\gamma - \gamma')$. This is so because the finite gap is infinitesimally small compared to the semi-infinite structure. When taking the correlations for (14), the integral arises

$$\int_{-\infty}^\infty \exp(-i(\gamma - \gamma')a)\tilde{F}(\gamma,\gamma')d\gamma d\gamma',$$

and for (15) there is also an integral in which there is a product $\gamma\gamma'$. The correlations in (14), (15) are the coefficients at delta-functions, and at the addition is a finite value at an infinitely large value of the delta function. For the *y*-components, similar relations are obtained.

For a compact representation of the crosslinking results, we introduce the notation $d_0 = \exp(-2ik_z^0 a) = \exp(-2\kappa a)$, $\tilde{d} = \exp(-2ik_z a)$, $C_+^0 = d_0 + 1$, $C_-^0 = d_0 - 1$, $C_+ = \tilde{d} + 1$, $C_- = \tilde{d} + 1$. We will crosslink the spectral fields at $z = a$ in the sequence:

$$E_{0x}^e + E_{0x}^h = E_x^e + E_x^h + E_x^0,$$

$$E_{0y}^e + E_{0y}^h = E_y^e + E_y^h + E_y^0,$$

$$H_{0x}^e + H_{0x}^h = H_x^e + H_x^h + H_x^0,$$

$$H_{0y}^e + H_{0y}^h = H_y^e + H_y^h + H_y^0$$

When stitching, the value $e_{\mathbf{q}}(\mathbf{r}_\tau)$ is reduced. Multiplying the third and fourth equations by $Z_0$, leaving the values $\left(E_x^0(\mathbf{q}), E_y^0(\mathbf{q}), Z_0 H_x^0(\mathbf{q}), Z_0 H_y^0(\mathbf{q})\right)$ on the right side of this system of equations and writing this 4-vector as $E^0(\mathbf{q})$, we transfer the remaining coefficients to the left and enter the 4-vectors

$$E_x(\mathbf{q}) = \left(E_{0x}^e(\mathbf{q}), E_x^e(\mathbf{q}), E_{0x}^h(\mathbf{q}), E_x^h(\mathbf{q})\right), \quad E_y(\mathbf{q}) = \left(E_{0y}^e(\mathbf{q}), E_y^e(\mathbf{q}), E_{0y}^h(\mathbf{q}), E_y^h(\mathbf{q})\right),$$

$$H_x(\mathbf{q}) = \left(H_{0x}^e(\mathbf{q}), H_x^e(\mathbf{q}), H_{0x}^h(\mathbf{q}), H_x^h(\mathbf{q})\right), \quad H_y(\mathbf{q}) = \left(H_{0y}^e(\mathbf{q}), H_y^e(\mathbf{q}), H_{0y}^h(\mathbf{q}), H_y^h(\mathbf{q})\right),$$

and $E_z(\mathbf{q}) = \left(E_{0z}(\mathbf{q}), E_z(\mathbf{q}), Z_0 H_{0z}(\mathbf{q}), Z_0 H_z(\mathbf{q})\right)$. As a result, we obtain two equivalent systems of four linear algebraic equations. The first one in the form $\hat{A}_{1m}(\mathbf{q}, d_0, \tilde{d}) E_{xm}(\mathbf{q}) = E_1^0(\mathbf{q})$, $\hat{A}_{2m}(\mathbf{q}, d_0, \tilde{d}) E_{ym}(\mathbf{q}) = E_2^0(\mathbf{q})$, $\hat{A}_{3m}(\mathbf{q}, d_0, \tilde{d}) H_{xm}(\mathbf{q}) = E_3^0(\mathbf{q})$, $\hat{A}_{4m}(\mathbf{q}, d_0, \tilde{d}) H_{ym}(\mathbf{q}) = E_4^0(\mathbf{q})$. Also the second system in the form $\hat{C}(\mathbf{q}, d_0, \tilde{d}) E_z(\mathbf{q}) = q^2 E^0(\mathbf{q})$. The equations of first system are

$$C_+^0 E_{0x}^e(\mathbf{q}) - C_+ E_x^e(\mathbf{q}) + C_+^0 E_{0x}^h(\mathbf{q}) - C_+ E_x^h(\mathbf{q}) = E_x^0(\mathbf{q}, a),$$

$$C_+^0 E_{0y}^e(\mathbf{q}) - E_y^e(\mathbf{q}) + C_+^0 E_{0y}^h(\mathbf{q}) - C_+ E_y^e(\mathbf{q}) = E_y^0(\mathbf{q}, a),$$

$$-\frac{C_-^0}{\rho_0^e} E_{0y}^e(\mathbf{q}) + \frac{C_-}{\rho^e} E_y^e(\mathbf{q}) - \frac{C_-^0}{\rho_0^h} E_{0y}^h(\mathbf{q}) + \frac{C_-}{\rho^h} E_y^h(\mathbf{q}) = Z_0 H_x^0(\mathbf{q}, a),$$

$$\frac{C_-^0}{\rho_0^e} E_{0x}^e(\mathbf{q}) - \frac{C_-}{\rho^e} E_x^e(\mathbf{q}) + \frac{C_-^0}{\rho_0^h} E_{0x}^h(\mathbf{q}) - \frac{C_-}{\rho^e} E_x^h(\mathbf{q}) = Z_0 H_y^0(\mathbf{q}, a).$$

The matrices of these systems have the form

$$\hat{A}(\mathbf{q}, d_0, \tilde{d}) = \begin{bmatrix} -C_+^0 & C_+ & -C_+^0 & C_+ \\ -C_+^0 & C_+ & C_+^0 & -C_+ \\ \dfrac{C_-^0}{\rho_0^e} & -\dfrac{C_-}{\rho^e} & -\dfrac{C_-^0}{\rho_0^h} & \dfrac{C_-}{\rho^h} \\ -\dfrac{C_-^0}{\rho_0^e} & \dfrac{C_-}{\rho^e} & -\dfrac{C_-^0}{\rho_0^h} & \dfrac{C_-}{\rho^h} \end{bmatrix}, \qquad (16)$$

$$\hat{C}(\mathbf{q}, d_0, \tilde{d}) = \begin{bmatrix} -C_+^0 k_z^0 k_x & C_+ k_z k_x & -C_+^0 k_0 k_y & C_+ k_0 k_y \\ -C_+^0 k_z^0 k_y & C_+ k_z k_y & C_+^0 k_0 k_x & -C_+ k_0 k_x \\ \dfrac{C_-^0}{\rho_0^e} k_z^0 k_y & -\dfrac{C_-}{\rho^e} k_z k_y & -\dfrac{C_-^0}{\rho_0^h} k_0 k_x & \dfrac{C_-}{\rho^h} k_0 k_x \\ -\dfrac{C_-^0}{\rho_0^e} k_z^0 k_x & \dfrac{C_-}{\rho^e} k_z k_x & -\dfrac{C_-^0}{\rho_0^h} k_0 k_y & \dfrac{C_-}{\rho^h} k_0 k_y \end{bmatrix}. \qquad (17)$$

The equivalence of the equations follows from the connection of the transverse components with the longitudinal ones. The elements of the matrix (17) have the dimension inverse to the square of the length, and the elements of the matrix (16) are dimensionless. They explicitly indicate the dependence on $a$ in the form of quantities $d_0$, $\tilde{d}$. The solution of the problem has the form $E_\tau(\mathbf{q},a) = \hat{B}(\mathbf{q},d_0,\tilde{d})E^0(\mathbf{q},a)$ or $E_z(\mathbf{q},a) = q^2\hat{D}(\mathbf{q},d_0,\tilde{d})E^0(\mathbf{q},a)$. Here we denote the inverse of (15) and (16) matrices $\hat{B} = \hat{A}^{-1}$, $\hat{D} = \hat{C}^{-1}$. To determine the dispersion force, we need correlations of the spatial components of the solution. To obtain them, we need to take the inverse Fourier transforms of the spatio-spectral components and construct correlations from them, i.e. multiply the spectral correlations by $(2\pi)^6 dk_x dk_y dk'_x dk'_y$ and integrate. When taking correlations, a delta function arises $(2\pi)^2 \delta(\mathbf{q}-\mathbf{q}')$, integration with which destroys one of the double spectral integrals. There is an integral $(2\pi)^4 dk_x dk_y$ with integrand functions $\langle E_n(\mathbf{q},a), E_n(\mathbf{q},a)\rangle$. We denote the coefficients for the delta function in correlations of type (13) as $\tilde{F}_n(\omega,T)$. For them $\tilde{F}_{1,2}(\omega,T) = Z_0^2 F(\omega,T)/(2k_0)^2$, $\tilde{F}_3(\omega,T) = (k_x^2/k_0^4)F(\omega,T)/4$, $\tilde{F}_4(\omega,T) = (k_y^2/k_0^4)F(\omega,T)/4$.

Then

$$\langle E_{zn}(\mathbf{q},a), E_{zn}(\mathbf{q},a)\rangle = Z_0^2 q^2 D_{nm}(\mathbf{q},d_0,\tilde{d})D^*_{nm}(\mathbf{q},d_0,\tilde{d})\tilde{F}_m(\omega,T). \tag{18}$$

Here $n = 1,2,3,4$, and over $m$ is the summation from 1 to 4.

We will determine the density of the dispersion force in terms of the energy-momentum tensor (EMT). The force acting on a body with a surface $S$ is defined in terms of the Maxwell tension tensor (MTT), which is equal to the minus sign of the spatial part of the EMT, which in the SI system has the form

$$T_{\alpha\beta} = \varepsilon_0 E_\alpha E_\beta + \mu_0 H_\alpha H_\beta - \delta_{\alpha\beta}[\varepsilon_0 \mathbf{E}^2 + \mu_0 \mathbf{H}^2]/2. \tag{18}$$

Greek indices run through the values of *x, y, z*, and the same indices are summed. EMT determines the pressure, and the tension force on the body is opposite and equal to

$$\mathbf{f} = -\int_S \hat{T}(\mathbf{r}_s)\mathbf{n}(\mathbf{r}_s)d^2r. \tag{19}$$

For random fields, take the autocorrelation function $\langle \hat{T}(\mathbf{r}_s)\rangle$. For real values, it means stochastic averaging. For complex amplitudes, the correlation function (hereinafter referred to as correlation) $\langle E_\alpha, E_\beta\rangle$ means the averaging of the value $E_\alpha E^*_\beta$. Since they need to be averaged over a period, a multiplier of 1/2 occurs. The left half-space is affected by a force along the *z*-axis

$$f_z = -\int_{-\infty}^{\infty}\int_{-\infty}^{\infty} \langle T_{zz}(x,y,-a)\rangle dxdy. \qquad (20)$$

Naturally, it is infinite, since, as can be seen, the integrand does not depend on the transverse coordinates. Therefore, we will consider it as a tension $T_z$:

$$T_x = \langle T_{xx}(x,y,-a)\rangle. \qquad (21)$$

Obviously, $T_x = \langle T_{xx}(x,y,-a)\rangle = -\langle T_{xx}(x,y,a)\rangle$, therefore, to determine the force density on the left half-space, we can use the results of crosslinking at $z=a$. In our case

$$\langle T_{zz}\rangle = \varepsilon_0 \frac{\langle E_z,E_z\rangle - \langle E_x,E_x\rangle - \langle E_y,E_y\rangle}{2} + \mu_0 \frac{\langle H_z,H_z\rangle - \langle H_x,H_x\rangle - \langle H_y,H_y\rangle}{2}. \qquad (22)$$

The tensor has the dimension of the energy density J/m$^3$ or N/m$^2$, so it determines the pressure (negative tension). We will use spectral values rather than time values. In harmonic fields, the dimension of the vectors is the same as in non-stationary ones. However, the spectral densities from time fields have a different dimension, the denominator of which includes the frequency. Accordingly, spectrum integration gives time fields and time correlations. The force will be determined by frequency integration, using spectral values. The values of the type $\varepsilon_0\langle E_x(\mathbf{r},\omega),E_y(\mathbf{r},\omega)\rangle$ have the dimension of action. In harmonic fields, you should use the values averaged over the period, i.e. in the correlation, and add a multiplier of 1/2.

Let us define the principles of finding the dispersion force. The force acting on the first body from the second should not depend on the field of the first body. It should tend to zero when the second body is removed to infinity. However, the field sources of the second body depend on fluctuations and induced currents in the first body through multiple diffraction. In our case, this is provided by waves of two directions in the gap with the correct choice of the phases of their decomposition coefficients, as well as a single spectral decomposition of fluctuation sources for both half-spaces. For half-spaces, the force is determined only by one of their boundaries. For the final samples, the force on the second far side should be taken into account. However, for an infinite left half-space, all fields radiated to the left by the right half-space are attenuated at an infinite length, which justifies the use of a single boundary. So, although we have determined the correlations of the components of all the transverse fields, then we will only need their correlations in the gap. But in addition to transverse correlations, you will also need longitudinal correlations. They can be expressed as follows:

$$\langle E_{0z}(\mathbf{q}),E_{0z}(\mathbf{q})\rangle = \frac{q^4 \langle E_{0x}^e(\mathbf{q}),E_{0x}^e(\mathbf{q})\rangle}{|k_z^0|^2 k_x^2} = \frac{q^4 \langle E_{0y}^e(\mathbf{q}),E_{0y}^e(\mathbf{q})\rangle}{|k_z^0|^2 k_y^2}, \qquad (23)$$

$$\langle H_{0z}(\mathbf{q}), H_{0z}(\mathbf{q}) \rangle = \frac{q^4 \langle E_{0x}^h(\mathbf{q}), E_{0x}^h(\mathbf{q}) \rangle}{Z_0^2 |k_z^0|^2 k_y^2} = \frac{q^4 \langle E_{0y}^h(\mathbf{q}), E_{0y}^h(\mathbf{q}) \rangle}{Z_0^2 |k_z^0|^2 k_x^2}. \tag{24}$$

Since all cross-correlations are expressed according to (16) in terms of matrix elements that depend on $d_0$ and $\tilde{d}$, then all correlations included in (22) depend on these values, i.e. ultimately on $a$. For example, we write $\langle E_{0x}^e(\mathbf{q}, d_0, \tilde{d}), E_{0x}^e(\mathbf{q}, d_0, \tilde{d}) \rangle$. This value, like other correlations, is included in the spectral value

$$\langle T_{zz}(\mathbf{q}, d_0, \tilde{d}) \rangle = \varepsilon_0 \frac{\langle E_z, E_z \rangle - \langle E_x, E_x \rangle - \langle E_y, E_y \rangle}{4} + \mu_0 \frac{\langle H_z, H_z \rangle - \langle H_x, H_x \rangle - \langle H_y, H_y \rangle}{4}, \tag{25}$$

corresponding to the space-time formula (22). In formula (25), all values depend on $(\mathbf{q}, d_0, \tilde{d})$ and are taken inside the gap, for example,

$$\langle E_x(\mathbf{q}, d_0, \tilde{d}), E_x(\mathbf{q}, d_0, \tilde{d}) \rangle = \langle (E_{0x}^e(\mathbf{q}, d_0, \tilde{d}) + E_{0x}^h(\mathbf{q}, d_0, \tilde{d}))(1 + d_0), (E_{0x}^e(\mathbf{q}, d_0, \tilde{d}) + E_{0x}^h(\mathbf{q}, d_0, \tilde{d}))(1 + d_0) \rangle =$$
$$(1 + d_0)^2 (\langle E_{0x}^e(\mathbf{q}, d_0, \tilde{d}), E_{0x}^e(\mathbf{q}, d_0, \tilde{d}) \rangle + \langle E_{0x}^h(\mathbf{q}, d_0, \tilde{d}), E_{0x}^h(\mathbf{q}, d_0, \tilde{d}) \rangle + 2\operatorname{Re}(\langle E_{0x}^e(\mathbf{q}, d_0, \tilde{d}), E_{0x}^h(\mathbf{q}, d_0, \tilde{d}) \rangle)).$$

The additional twos in the denominator correspond to the averaging over the period. When $a \to \infty$ we have $d_0 = \exp(-a\sqrt{q^2 - k_0^2}) \to 0$ and $\tilde{d} = \exp(-a\sqrt{q^2 - \varepsilon k_0^2}) \to 0$. The latter relation follows from the losses in the dielectric, and the former at $q^2 > k_0^2$. If $q^2 < k_0^2$, we have $|d_0| = 1$, $|\tilde{d}| = 1$, so this area should be excluded from the integration as not bringing the result into force. To exclude self-action, i.e., a term that does not depend on the distance, we should take $\langle T_{zz}(\omega, \mathbf{q}, a) \rangle^\sim = \langle T_{zz}(\omega, \mathbf{q}, d_0, \tilde{d}) \rangle - \langle T_{zz}(\omega, \mathbf{q}, 0, 0) \rangle$. This result means that the body itself does not act on itself. More precisely, the impact can be in the form of internal ponderomotive forces. Here we have explicitly indicated the frequency dependence. Finally, we have (22) in the form

$$\langle T_{zz} \rangle = \frac{1}{\pi^3} \int_0^\infty \iint_{q^2 > k_0^2} \langle T_{zz}(\omega, \mathbf{q}, a) \rangle^\sim d^2q\, d\omega. \tag{26}$$

The components of the matrix (16) are even in $k_x$ and $k_y$. As it is easy to see, the solution based on the matrix (17) is also even in $k_x$ and $k_y$. The result (26) is obtained by switching to positive frequencies and integrating over positive components and with the volume element in q-space $d^2q = dk_x dk_y$. Therefore, the result is multiplied by 8. We write $\langle T_{zz}(\omega, \mathbf{q}, a) \rangle^\sim = \varepsilon''(\omega) \Theta(\omega, T) \langle \tilde{T}_{zz}(\mathbf{q}, a) \rangle^\sim / \pi$, that the integral (26) takes the form

$$\langle T_{zz} \rangle = \frac{1}{\pi^4} \int_0^\infty \omega \varepsilon''(\omega) \Theta(\omega, T) \int_{k_0}^\infty \int_0^{\pi/2} \langle \tilde{T}_{zz}(\mathbf{q}, a) \rangle^\sim q\, dq\, d\vartheta\, d\omega. \tag{27}$$

In this integral, the variables are replaced as $k_x = q\cos(\vartheta)$, $k_y = q\sin(\vartheta)$. The angle integration can be performed approximately by the mean value theorem, assuming $\vartheta = \pi/4$ and multiplying by $\pi/2$. In (27), the substitution $k_x^2 = q^2/2$, $k_y^2 = q/2$, should be made, since in (27) there are only even powers of these components. Finally we have

$$\langle T_{zz} \rangle = \frac{1}{2\pi^3} \int_0^\infty \omega \varepsilon''(\omega) \Theta(\omega, T) \int_{k_0}^\infty \langle \tilde{T}_{zz}(\mathbf{q}, a) \rangle^\sim q \, dq \, d\omega. \tag{28}$$

In this expression

$$\langle E_{0z}(\mathbf{q}, a), E_{0z}(\mathbf{q}, a) \rangle = Z_0^2 q^2 D_{1m}(\mathbf{q}, d_0, \tilde{d}) D_{1m}^*(\mathbf{q}, d_0, \tilde{d}) \tilde{F}_m(\omega, T),$$

$$\langle H_{0z}(\mathbf{q}, a), H_{0z}(\mathbf{q}, a) \rangle = q^2 D_{3m}(\mathbf{q}, d_0, \tilde{d}) D_{3m}^*(\mathbf{q}, d_0, \tilde{d}) \tilde{F}_m(\omega, T),$$

$$\langle E_{0x}^e(\mathbf{q}, a), E_{0x}^e(\mathbf{q}, a) \rangle = (k_x k_z^0)^2 \langle E_{0z}(\mathbf{q}, a), E_{0z}(\mathbf{q}, a) \rangle / q^4,$$

$$\langle E_{0y}^e(\mathbf{q}, a), E_{0y}^e(\mathbf{q}, a) \rangle = (k_y k_z^0)^2 \langle E_{0z}(\mathbf{q}, a), E_{0z}(\mathbf{q}, a) \rangle / q^4,$$

$$\langle E_{0x}^h(\mathbf{q}, a), E_{0x}^h(\mathbf{q}, a) \rangle = (k_y k_0)^2 \langle H_{0z}(\mathbf{q}, a), H_{0z}(\mathbf{q}, a) \rangle / q^4, \tag{30}$$

$$\langle E_{0y}^h(\mathbf{q}, a), E_{0y}^h(\mathbf{q}, a) \rangle = (k_x k_0)^2 \langle H_{0z}(\mathbf{q}, a), H_{0z}(\mathbf{q}, a) \rangle / q^4,$$

$$\langle H_{0x}^e(\mathbf{q}, a), H_{0x}^e(\mathbf{q}, a) \rangle = (k_y k_0)^2 \langle E_{0z}(\mathbf{q}, a), E_{0z}(\mathbf{q}, a) \rangle / q^4,$$

$$\langle H_{0y}^e(\mathbf{q}, a), H_{0y}^e(\mathbf{q}, a) \rangle = (k_x k_0)^2 \langle E_{0z}(\mathbf{q}, a), E_{0z}(\mathbf{q}, a) \rangle / q^4,$$

$$\langle H_{0x}^h(\mathbf{q}, a), H_{0x}^h(\mathbf{q}, a) \rangle = (k_x k_z^0)^2 \langle H_{0z}(\mathbf{q}, a), H_{0z}(\mathbf{q}, a) \rangle / q^4,$$

$$\langle H_{0y}^h(\mathbf{q}, a), H_{0y}^h(\mathbf{q}, a) \rangle = (k_y k_z^0)^2 \langle H_{0z}(\mathbf{q}, a), H_{0z}(\mathbf{q}, a) \rangle / q^4.$$

Expressing the cross-correlations in terms of the longitudinal ones and taking into account that the longitudinal ones are proportional to $F_n$, and from here, to $\tilde{F}$, we get expressions $\langle \tilde{T}_{zz}(\mathbf{q}, a) \rangle$ in terms of the field correlations. From the correlations (30), we subtract their values for $d_0 = 0$ and $\tilde{d} = 0$, and get $\langle \tilde{T}_{zz}(\mathbf{q}, a) \rangle^\sim$. The value $\langle \tilde{T}_{zz}(\mathbf{q}, a) \rangle^\sim$ depends on the frequency. At high frequencies the $\det(\hat{C}(\mathbf{q}, d_0, \tilde{d}))$ it is proportional to $k_0^2$, so the matrix elements in $\hat{D}(\mathbf{q}, d_0, \tilde{d})$ either decrease or do not increase. Due to the presence of multipliers $d_0, \tilde{d}$, the internal integral in (28) converges. However, it can even be limited to a certain value $q_{msx}$. This value is related to the actual discreteness of the medium. Therefore, when $k_0 = k_{max}$ the integral vanishes. If we assume that $\hat{D}(\mathbf{q}, d_0, \tilde{d})$ does not decrease, then for the convergence of the frequency integral, it is necessary to fulfill the inequality $\varepsilon''(\omega)\Theta(\omega, T) < K/\omega$ at high frequencies, where K is a constant. Since $\varepsilon''(\omega)\Theta(\omega, T) \approx \omega\hbar/2$, then for the Lorentz dispersion holds $\varepsilon''(\omega)\Theta(\omega,T) \to K/\omega$. For her as well, so there are no problems at low frequencies. For convergence, the internal integral must decrease with frequency. With the Drude dispersion for the conducting half-planes, the poles are at zero, and the divergence of the frequency integral is at zero. This divergence is removed if we use the Drude-Smith model. Using the model

$$\langle T_{zz}\rangle = \frac{1}{2\pi^3}\int_0^{cq_{max}}\omega\varepsilon''(\omega)\Theta(\omega,T)\int_{k_0}^{q_{max}}\langle\tilde{T}_{zz}(\mathbf{q},a)\rangle^{\sim} q\,dq\,d\omega, \qquad (31)$$

it is quite possible to calculate the specific force.

In [8], a formula is given in which the contributions of TM-modes (E-modes) and TM-modes (H-modes) are separated. In our case, the problem is solved strictly, but there is no separation, since these modes are tied through fluctuation fields that are not separated into E-modes and H-modes. In contrast to [8], integration can be performed along the real axis of the vector waveguide module and along the real frequency axis. For a rigorous calculation, it is necessary to perform numerical integration with respect to the angle in the space of wave vectors. In order for the division into TM and TE modes to occur, it is necessary that the matrix $\hat{C}$ can be converted to a block-diagonal, which is not the case. You can use another classical electrodynamic approach to determining the tension, based not on the decomposition and crosslinking of fields, but on the representation of fields through Green's functions. In this case, the total vector potential $\mathbf{A} = \mathbf{A}^0 + \mathbf{A}^d$ is the sum of the fluctuation and diffraction parts. The diffraction part is determined by the induced (diffraction) currents $\mathbf{J}(\mathbf{r})$, and the fluctuation part $\mathbf{J}^0(\mathbf{r})$ which is determined from the fluctuation-dissipation theorem. For this solution of the problem, it is convenient to take the thickness $t$ of the finite one, and then go to the limit. The spectral electric and magnetic fields in the structure under consideration are defined as [21, 22] (the dependence on $\omega$ is omitted)

$$\mathbf{E}(\mathbf{r}) = \mathbf{E}^0(\mathbf{r}) + \int_{V_1+V_2}\hat{\Gamma}^e(\mathbf{r},\mathbf{r}')\mathbf{J}(\mathbf{r}')d^3r', \qquad (32)$$

$$\mathbf{H}(\mathbf{r}) = \mathbf{H}^0(\mathbf{r}) + \int_{V_1+V_2}\hat{\Gamma}^h(\mathbf{r},\mathbf{r}')\mathbf{J}(\mathbf{r}')d^3r'. \qquad (33)$$

Here, the index "zero" denotes the fluctuation fields excited by current fluctuations:

$$\mathbf{E}^0(\mathbf{r}) = \int_{V_1+V_2}\hat{\Gamma}^e(\mathbf{r},\mathbf{r}')\mathbf{J}^0(\mathbf{r}')d^3r', \qquad (34)$$

$$\mathbf{H}^0(\mathbf{r}) = \int_{V_1+V_2}\hat{\Gamma}^h(\mathbf{r},\mathbf{r}')\mathbf{J}^0(\mathbf{r}')d^3r'. \qquad (35)$$

Volumes $V_n$ characterize the infinite in the limit of the plate area. These fields in turn excite the diffraction fields and create full fields. The complete fields satisfy the impedance conditions of the form $\mathbf{E}_n = i\omega\varepsilon_0(\varepsilon-1)\mathbf{J}_n$. Here, the index refers to each of the bodies (the observation point belongs to body $n$). Tensor GF are defined in terms of the scalar GF [23, 24]:

$$\hat{\Gamma}^e(\mathbf{r},\mathbf{r}') = \frac{k_0^2 \hat{I} + \nabla \otimes \nabla}{i\omega\varepsilon_0} G(\mathbf{r}-\mathbf{r}'),$$

$$\hat{\Gamma}^h(\mathbf{r},\mathbf{r}') = \hat{R} G(\mathbf{r}-\mathbf{r}'),$$

where the tensor $\nabla \otimes \nabla$ has components, and the matrix $\hat{R}$ corresponds to the operator $\nabla \times$ ("curl") and has the form

$$\hat{R} = -\hat{\varepsilon}_{\alpha\beta\gamma}\partial_\gamma = \begin{bmatrix} 0 & -\partial_z & \partial_y \\ \partial_z & 0 & -\partial_x \\ -\partial_y & \partial_x & 0 \end{bmatrix}.$$

Here a completely antisymmetric Levi-Civitta tensor $\hat{\varepsilon}_{\alpha\beta\gamma}$ is introduced. We have the components $\Gamma^e_{\mu\nu} = (i\omega\varepsilon_0)^{-1}(\delta_{\mu\nu} + \partial_\mu\partial_\nu)G$, $\Gamma^h_{\mu\nu} = -\varepsilon_{\mu\nu\kappa}\partial_\kappa G$. It is convenient to take the spectral representation of GF in the form [21, 22]

$$G(\mathbf{r}) = \int_{-\infty}^{\infty}\int_{-\infty}^{\infty} \frac{\exp(-i\alpha x - i\beta y - \kappa|x|)}{8\pi^2 \kappa} d\gamma d\beta. \tag{36}$$

In it $\kappa = \sqrt{\alpha^2 + \beta^2 - k_0^2}$, and in the case of $q^2 = \alpha^2 + \beta^2 < k_0^2$ we taking a branch of the root $\kappa = i\sqrt{k_0^2 - q^2}$. In this case, these are waves radiated along the $x$-axis in both directions. We introduce the vectors $\mathbf{k} = (\alpha,\beta,\gamma)$, $\mathbf{q} = (\alpha,\beta)$. For spectral equations, it is convenient to use the following spectral representation of GF:

$$G(\mathbf{r}) = \int_{-\infty}^{\infty}\int_{-\infty}^{\infty}\int_{-\infty}^{\infty} \frac{\exp(-i\mathbf{k}\mathbf{r})}{8\pi^3(\mathbf{k}^2 - k_0^2)} d^3k. \tag{37}$$

Let denote $\Gamma^e_{\mu\nu}(\mathbf{k}) = (i\omega\varepsilon_0)^{-1} g_{\mu\nu} G$. The spectral representations of the tensor GF are obtained by replacing, for example, $\partial_z \to -i\gamma$, and so on. Passing to the spectral equations, we obtain an algebraic system of equations. It should be noted that it is impossible to strictly use EMT in this approach, since it is necessary to take into account the forces acting from the inner sides of the plate, the components of EMT in which are not known. You can use an approach based on the Lorentz force. The electric part of this force $-\omega\varepsilon_0 \mathrm{Re}((\nabla\cdot(\varepsilon-1)\mathbf{E})\mathbf{E}^*)/2$ acts at the interface (where there is an induced surface charge density associated with the DP jump), and the magnetic part is equal to $\omega\mathrm{Re}(i(\varepsilon-1)\mathbf{E}\times\mathbf{H}^*)/(2c^2)$. This volumetric force should be integrated over the thickness and find the limit. Let us denote $\mathbf{J} = i\omega\varepsilon_0(\varepsilon-1)\mathbf{E} = \xi\mathbf{E}$, $\xi = Z_0\omega\varepsilon_0$, go to the spectral quantities, represent the electric field in each of the layers in terms of GF (36), and impose the condition $\mathbf{J} = \xi \mathbf{E}$. We get a system of equations

$$J_{1x}(\eta - \xi g_{xx}) - J_{1y}\xi g_{xy} - (J_{2x}\xi g_{xx} + J_{2y}\xi g_{xy})d_0 =$$
$$= J^0_{1x}\xi g_{xx} + J^0_{1y}\xi g_{xy} + (J^0_{2x}\xi g_{xx} + J^0_{2y}\xi g_{xy})d_0 \quad,$$

$$-J_{1x}\xi g_{xy} + J_{1y}(\eta - \xi g_{yy}) - (J_{2x}\xi g_{xy} + J_{2y}\xi g_{yy})d_0 =$$
$$= J^0_{1x}\xi g_{xy} + J^0_{1y}\xi g_{yy} + (J^0_{2x}\xi g_{xy} + J^0_{2y}\xi g_{yy})d_0 \quad,$$

$$-(J_{1x}\xi g_{xx} + J_{1y}\xi g_{xy})d_0 + J_{2x}\xi(\eta - \xi g_{xx}) - J_{2y}\xi g_{xy} =$$
$$= (J^0_{1x}\xi g_{xx} + J^0_{1y}\xi g_{xy})d_0 + J^0_{2x}\xi g_{xx} + J^0_{2y}\xi g_{xy} \quad,$$

$$-(J_{1x}\xi g_{xy} + J_{1y}\xi g_{yy})d_0 - J_{2x}\xi g_{xy} + J_{2y}\xi(\eta - \xi g_{yy}) =$$
$$= (J^0_{1x}\xi g_{xy} + J^0_{1y}\xi g_{yy})d_0 + J^0_{2x}\xi g_{xy} + J^0_{2y}g_{yy} \quad.$$

They are marked $\eta = 2ik_0\kappa$, and the indexes 1 and 2 correspond to the first and second plates. Denoting the 4-vectors $\bar{J} = (J_{1x}, J_{1y}, J_{2x}, J_{2y})$, $\bar{J}^0 = (J^0_{1x}, J^0_{1y}, J^0_{2x}, J^0_{2y})$, we represent this system equation in the form $\bar{\bar{A}}(\eta, d_0)\bar{J} = -\bar{\bar{A}}(0, d_0)\bar{J}^0 = \tilde{J}^0$ or $\bar{J} = \bar{\bar{A}}^{-1}(\eta, d_0)\tilde{J}^0$. The matrix is indicated here

$$\bar{\bar{A}}(\eta, d_0) = \begin{bmatrix} \eta - \xi g_{xx} & -\xi g_{xy} & -\xi g_{xx}d_0 & -\xi g_{xy}d_0 \\ -\xi g_{xy} & \eta - \xi g_{yy} & -\xi g_{xy}d_0 & -\xi g_{yy}d_0 \\ -\xi g_{xx}d_0 & -\xi g_{xy}d_0 & \eta - \xi g_{xx} & -\xi g_{xy} \\ -\xi g_{xy}d_0 & -\xi g_{yy}d_0 & -\xi g_{xy} & \eta - \xi g_{yy} \end{bmatrix}. \tag{38}$$

The solution of the problem, as above, is reduced to the inversion of the matrix . By inverting the matrix, you can calculate correlations and define fields. In this approach, approximate solutions can be found, since the system of equations has the form of Lippmann-Schwinger equations. However, the approximate approach requires the calculation of multiple integrals in the space of wave vectors. Also, as above, the terms with $d_0 = 0$ should be excluded form the all terms. The results for Lorenz dispersion force model for two dielectric plates are presented in Fig. 2.

The author is grateful to Dr. I. S. Nefedov for discussing the results. The work was supported by the Ministry of Education and Science of the Russian Federation as part of the state task (project No. FSRR-2020-0004).

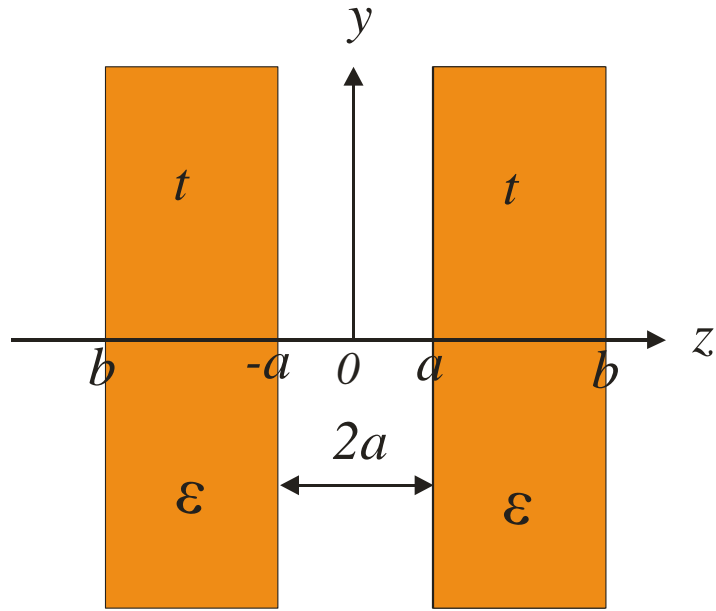

Figure 1. Configuration in the form of two infinite in *x* and *y* dielectric plates in a vacuum

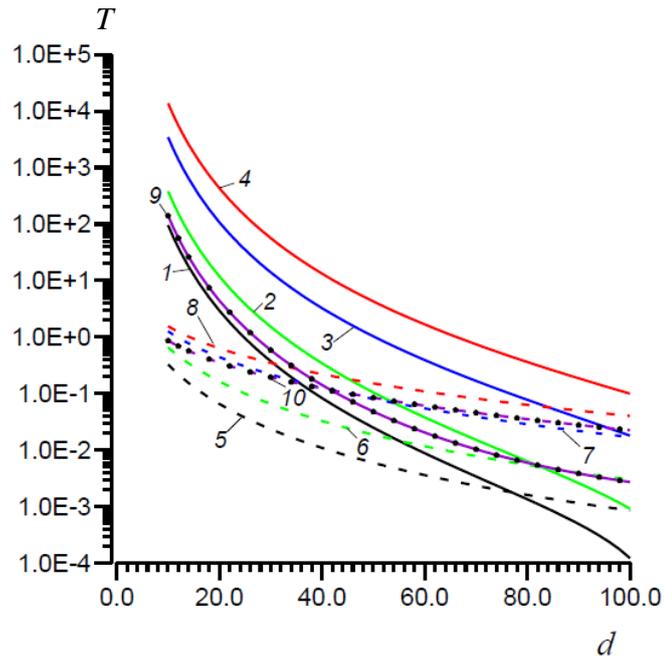

Figure 2. Surface force densities (in N/m$^2$) $T$: (curves 1, 2, 3, 4), (dashed curves 5, 6, 7, 8, 10) and (9) for the interaction of two identical plates of thickness $t$ as a function of the distance $d$ (nm) at T=300 K. Curves 1, 5 correspond to $t$=5 nm, 2, 6 $t$=10 nm, 3, 7 $t$=30 nm, 4, 8–10 $t$=60 nm. Curves 1-8 correspond to metal plates, curves 9,10 correspond to dielectric plates. The metal is modeled according to the Drude-Lorenz-Smith model with the parameters $\varepsilon_L = 10$, $\omega_p = \Omega_p = 1.6 \cdot 10^{16}$, $\Omega_0 = 1.8 \cdot 10^{12}$, $\Omega_c = 2.4 \cdot 10^{13}$, $\omega_0 = 3.0 \cdot 10^{16}$, $\omega_c = 1.8 \cdot 10^{12}$ (frequencies in Hz). For the dielectric model $\Omega_0 = 1.8 \cdot 10^{13}$, $\omega_c = 1.8 \cdot 10^{14}$